\documentclass[12pt,preprint]{aastex}
\usepackage{graphicx}

\begin{document}

\title{\bf X-ray Bright QSO's around NGC 3079}

\author{H. Arp}
\affil{Max-Planck-Institut f\"ur Astrophysik, Karl Schwarzschild-Str.1,
  Postfach 1317, D-85741 Garching, Germany}
 \email{arp@mpa-garching.mpg.de}

\author{E. M. Burbidge}
\affil{Center for Astrophysics and Space Sciences
0424, University of California, San Diego, CA 92093-0424, USA}

\begin{abstract}
NGC 3079 is a very active, disturbed galaxy which has been observed to have X-ray 
and radio ejections from it as well as an optical superbubble along its minor axis. Here we show that 
the brightest X-ray sources within about 40 arcmin are in large excess of background values. 
The X-ray sources are identified as quasars and AGN's which are aligned and spaced across the 
Seyfert nucleus to a degree which is unlikely to be due to chance. The famous double 
quasar which has been interpreted as a gravitational lens is discussed in terms of the the X - ray 
and ULX sources which appear associated with NGC 3079.
\end{abstract}


\section{INTRODUCTION}

Among bright galaxies N3079 is one of the most optically disturbed. It is listed in the Revised Shapley 
Ames (Sandage and Tamman 1981) as $B_T$ = 11.20 mag., or inclination corrected as 10.31 mag.  
Previous observations have led to the conclusion that there is a super wind 
from the nucleus and gas flow between the disk and the halo. The relevant parameters are discussed 
in detail and referenced in Pietsch et al. (1998). For the purposes of the present paper the most 
important properties are the active nucleus (LINER/Seyfert 2), radio and X-ray material ejected that 
nucleus and X-ray and radio material extending into a region around NGC 3079.

In the present investigation we are first concerned with the density of strong X-ray sources in the
immediate vicinity NGC 3079. We find a strong over density of bright X-ray sources compared to
conservative values of expected background sources. This result is supported by the earlier
studies of Radecke (1997) and Arp (1997), who showed that there is a strong tendency for
X-ray sources to cluster about active Seyfert galaxies. Almost all the sources then turn out to be high
redshift QSO's and AGN's. Many cases of this kind have been found
(eg. near the AGN galaxies NGC 1068, 2639, 3516, 3628).  (For review see Arp 1998; 2003.) The 
typical separations between the QSOs and the galaxies in those cases are d'  $\sim$ 15\arcmin 
to 20\arcmin.  It is clear that if the separations are smaller than this, as is the case of NGC 3079 
there will be an even greater likelihood that the QSOs and galaxies are physically associated. 
It is also noted that the brightness of the X-ray sources near NGC 3079 place them in the
category of Ultra Luminous X-ray sources (ULX's) - objects which have heretofore been accepted
as belonging to their nearby galaxies (Colbert, E. \& Ptak A. 2002; Burbidge et al. 2003; Arp et el. 2004). 

Having established the general association of the nearby quasars with NGC 3079 we then explore their 
possible interaction with the Seyfert's ejected X-ray and radio material.

\section{THE BRIGHTEST X-RAY SOURCES AROUND NGC 3079}

In Fig. 1 we show that there are four extremely strong X-ray sources within 36 arcmin of NGC 3079.
All are optically identified as AGN/QSO's and Table 1 lists their observed properties. 
1RXS J095834.2+560228 was observed by E. M. Burbidge and H. Arp at the Lick 3.3  meter telescope 
on Mt. Hamilton. They derived the listed redshift of z = .2154. They also derived a redshift of z = 1.024
for SBS 0955+560. A redshift for this same quasar of z = 1.021 is listed in NED and was obtained by 
Stepanian et al. in 1990. Finally the bright AGN 33' to the SE of NGC 3079 was measured by 
C. Guti\'errez with the INT in Las Palmas. A redshift of z = .431 was derived for this very radio bright 
and, very interestingly, a barely resolved triple radio source (FIRST). 

The extremely bright double quasar usually interpreted as a gravitational lens has listed redshifts of z =
1.413 and 1.415.

 A glance at Fig 1 appears show significant alignment across the center of NGC 3079. Therefore we 
first test the mathematical probability that this occurence could be accidental: The lowest flux of the 
four is  
$ F = 64 x 10^{-14}ergcm^{-2} sec{^-1}$. Using the RIXOS  logN-logS counts of  bright AGN's over 
the sky (Page et al. 1996, Fig.3) we obtain a value of $N = .11 deg^{-2}$ for this bright or brighter flux.
Within the 36' radius of the four in Fig. 1 there is $1.1 {deg^2}$. Therefore we expect 0.12 AGN's this
bright in an area where we find four. By chance this is a probability about $2 x 10^{-4}$. Or if we take
$\sqrt0.12 = sigma =.35$ then we have about a 10 sigma deviation from expected numbers. 

\begin{table}
\caption{Bright X-ray Quasars around NGC 3079 ($F_X \geq 50, d' \leq 36'$)} 
\label{Table1} \vspace{0.3cm}
\begin{tabular}{lcccl}
Quasar & R mag. & $F_X $ & z & Remarks\\
& & \\
1RXS J095834.2+560228 & 17.3 & 89 & .215 & ft. nvss, FiRST, 2MASX\\
SBS 0955+560 & 17.5 & 64  & 1.024 & 2MASX\\
87GB 100156.9+553816 & 19.3 & 65 & .431 & strong nvss, triple FiRST\\
Q0957+561A,B & 14.9/15.4 & 201 & 1.415 & "lensed" dbl QSO\\
\end{tabular}
\end{table}

\subsection{The famous double, "lensed" quasar}

By far the brightest and closest to NGC 3079 is the double quasar 0957+561A,B at F= 200 and d' =
14' distance. A very conservative (only X-ray sources) estimate of the density of such high flux  
X-ray sources is $N = .03 deg^{-2}$ (Hasinger et al. 1993). The area enclosed by 14' is 
$0.17 deg^{-2}$ so the chance of accidental nearness is about $P = 5 x 10^{-3}$. Actually we will 
see later in the discussion of RIXOS background densities that this probability should be much reduced.

NGC 3079 and 0957+561A,B by themselves are such unusual examples of ejecting galaxies and 
quasars that it is interesting to estimate the chances of their close association by several different 
methods. Since there are only about 186 galaxies with $B_T$ = 11.20 mag. or brighter than NGC 3079 
in the sky (See E.M. Burbidge et al. 2004) the chance of finding one within 14' of the very bright, 
highly unusual quasar is only about $P = 7.7 x10^{-4}$. One so disturbed would be about an order of 
magnitude less. If we take 0957+561A,B as on of the two brightest "lensed" quasars in the sky, then 
$P = 8 x 10^{-6}$ of finding it so close to NGC 3079. In a later section we will investigate the question
of whether there is evidence for a physical connection to NGC 3079.

\section{THE BRIGHTEST X-RAY SOURCES WITHIN 15' OF NGC 3079}

Fig. 2 shows the X-ray sources within 15' of NGC 3079 which have Flux greater than 10 (Pietsch et 
al. 1998).  The density of X-ray sources as great or greater than this value is N = $3.2 deg^{-2}$ 
from Fig. 3 of (Hasinger et al. 1998). Within a radius of 15' then we would expect N = 0.63 sources. 
From our figure 2 we see there are 6 such sources. This would translate into a probability of 
P = $6 x 10^{-2}$. If, however, we take the density of X-ray AGN's rather than just X-ray sources 
we can read N = $1.8 deg^{-2}$ from the RIXOS survey as portrayed in Fig. 3 of Page et al. 1996. 
Within 15' radius of Fig.3 then we would expect only .35 X-ray AGN's whereas we find 6. This would 
give $2 x 10^{-3}$ chance probability.Or if we take $\sqrt0.35= sigma =.59$ then we have about 
a 9.6 sigma result. 

This last expectation probability, however, depends on the RIXOS representing the background
density of  X-ray AGN's away from active galaxies. In fact, however, the RIXOS (and other
AGN surveys) were made from fields whose target centers were to a large extent active galaxies. 
Using sky samples far from active galaxies would reduce the probabilties we have just calculated by 
large factors, making their association with NGC 3079 even more certain. 

\begin{table}[h]
\caption{Bright X-ray Quasars around NGC 3079 ($F_X \geq 10, d' \leq 15'$)} 
\label{Table2} \vspace{0.3cm}
\begin{tabular}{lcccl}
Quasar & R mag. & $F_X $ & z & Remarks\\
Q0957+561A,B & 14.9/15.4 & 201 & 1.414 & "lensed" dbl QSO\\
RX J100119.7+554559 & 19.5& 10 & .679\\
RX J100056.4+534100 & 19.6 & 19 & 1.038 & strong nvss, slipped\\
RX J100032.3+553631 & 17.4 & 43 & .216 & Seyfert 2\\
RX J100309.6+554135 & 18.9 & 15 & .673 & wk. nvss rad source\\
RX J100110.0+552838 & 17.2 & 20 & 2.091 & wk 2MASX source
\end{tabular}
\end{table}

\section{THE PROBLEM WITH RIXOS}

The RIXOS survey was made from fields that had been observed by pointing at interesting
targets. The targets were generally strong X-ray sources such as Seyfert galaxies or low redshift 
AGN's. 

Quoting from Page et al. (1996):
"The target of each observation and a small region around it has been excluded from the analysis, so
that RIXOS consists entirely of serendipitously discovered sources. Sources more than 17 arcmin
off-axis have also been excluded . . ." ". . . spectra were taken for all optical counterparts within the 
1 $sigma$ error circle of each X-ray source."

The redshifts of the measured RIXOS sources were much higher than the central target and were
therefore assumed to be background objects at much greater distances. This assumption has not been
questioned in spite of the analysis by H.D. Radecke (1996; 1997) that demonstrated a 7.5 sigma 
association between bright Seyfert galaxies and bright X-ray sources within about 10' to 25'. Many of
these were identified as high redshift quasars (Arp 1997).

To illustrate the problems which this causes for the standard background density of bright X-ray
AGN's we cite only three cases:

1) Around NGC 3079 there are 6 RIXOS sources which we have just seen have a very high probability
of belonging to the active central galaxy. If this is true, then instead of having 6 RIXOS background
sources we have 0!

2) In the NGC 3516 field there are 5 RIXOS quasars. They are part of a line coming out along the minor 
axis of this extremely active Seyfert galaxy (Chu et al. 1998;  Arp 1999). They are called by some the 
Rosetta stone for ejection of quasars. If we associate them with NGC 3516 the number of 
RIXOS background AGN's goes from 5 to 0!

3) But even In a field centered on a very bright white dwarf we find off to the NW an Atlas peculiar
galaxy with an X-ray jet emerging from it. Closely surrounding this active galaxy are three RIXOS 
quasars of very closely the same redshift  (Pietsch and Arp 2001)  So even in a field pointed at a 
galactic star we find that 3 RIXOS quasars which are listed as background, go to 0!

It would be necessary to now identify and review each of the 82 fields used by RIXOS in order to get 
a realistic background count of bright X-ray AGN's. Whatever that turns out to be, however, it is
clearly much lower than we used in the above association probabilities with NGC 3079. It  therefore
strengthens immensely the evidence for the physical association of the bright quasars around the
ejecting, disturbed Seyfert NGC 3079 which has been found in the present paper.

\section{PAIRING CHARACTERISTICS OF THE NGC 3079 QUASARS}

Previous associations of X-ray quasars with ejecting galaxies have shown a tendency to occur in
evenly spaced pairs with lower redshift quasars being further out and higher redshift  quasars being
less separated from the galaxy (Arp 1999). In Fig. 1 the z = .215 and 1.021 quasars 
are both rather bright optically. They are fairly well aligned and evenly spaced across NGC 3079 with 
the blazar/AGN of z = .431 to the SE. 
The z = .431 and .215 objects make the best pair and could be explained by by an ejection of z = .30 
quasars at velocity $z_v = +.10$  and  $-.07$. (z = .30 being a major quantized intrinsic redshift peak.) 
If we compute the probability of the z = .215 and .431 quasars being this close to NGC 3079 and being 
aligned across it within about 4 deg. and equally spaced to about 15\%, then we have a probability of
accidently having this previously established configuration of about P =$1 x 10^{-4}$.

While the z = .215 quasar is a very strong X-ray source it is only a weak radio and 2 mm infra red 
source. On the other hand the z = .431 source is fainter optically but a very strong radio and 2MASX.
source. The high resolution FIRST map shows that it is a very compact triple source 
(like the triple source/quasar along the minor axis of NGC 2639 as shown in  E.M. Burbidge et al
(2004). Being strong in both radio and X-ray suggests it is in the BL Lac class, which kind of quasars
were found to be strongly represented among ULX's around parent galaxies (Arp et al. 2004).
\footnote{36" S of the X-ray position, and 17" S of the z = .431 AGN, we found a quasar of z = .326. It
is about the same magnitude, r = 19.1 b = 19.2 and may be associated with the z = .431 object. It
makes a better pair with the z = .215 quasar around the z = .30 Karlsson peak redshift.}

At lesser but still strong X-ray fluxes shown in Fig. 2 the most conspicuous pair is the z = .673 and 
1.038 quasars aligned closely along the minor
axis of NGC 3079. They are at d' = 10.1 and 8.7 respectively. This would give a probability of .15 x .15
of the two being found so close to NGC 3079. An alignment of 8/90 deg. and a spacing equal to about 
15\% would give a probability of P = $3 x 10^{-4}$ for this a priori configuration.

Of course there are 4 other quasars in this area which increases the probability just computed. 
However there are also further considerations supporting the relation of the aligned pair. First 
there is the fact that the pair is coming out closely along the minor axis where it is observed that 
actual ejection of radio and X-ray material is taking place. Fig. 4 here shows the PSPC X-ray 
isophoteds streaming away to the W, close to the extension of the minor axis. Fig. 3 here shows the 
high resolution radio lobes coming out along the minor axis. There does not seem much doubt that 
something is being ejected in these directions. The z = 1.038 quasar seems to be the obvious
candidate. Moreover in the case of NGC 3079, the strong 
disruption exhibited by the optical image makes it seem likely that there has also been ejection in 
other directions than the minor axis. Since the surroundings of this extremely active galaxy must be 
considerably crowded and denser than usual it seems necessary to consider the effect on the ejecta 
by this denser medium surrounding the galaxy.

\subsection{Quantization of Redshifts}

The uniquely famous pair of bright optical and X-ray quasars known as 0957+562A,B are only about 
14' NNE of NGC 3079. As noted before it would be rather appropriate if such an object fell close to
one of the brightest, most obviously ejecting galaxies in the sky. As for its redshift: In the reference
frame of NGC 3079 the quasars would have redshifts of (1 + 1.414)/(1 + .0038) = 1 + 1.405). This 
should be compared to the major quantized Karlsson redshift peak of $z_p = 1.41$, a difference of only
.005.

The remaining quasars are about 3\%  higher than the normal predicted Karlsson peaks. However they 
fit very well the shifted Karlsson values found for radio quasars in the R.A. = 12 hour region of the 
sky (see Arp et al. 1990, Fig. 3a and b). E.g. :

$z = .679 = z_0 = .673$
  
\centerline                             {re Karlsson $z_0 = .65 ~~~~~~~(12^h$ direction)}

$z = .673 = z_0 = .667$

$z = 1.021 = z_0 = 1.013$

\centerline                                    {Karlsson $z_0 = 1.02 ~~~~~~~(12^h$ direction)}           

$z = 1.038 = z_0 = 1.030$
 
\vskip0.8cm

$z = 2.091 = z_0 = 2.079$~~~~~~~{Karlsson $ z_0 = 2.05  ~~~~~~~(12^h$ direction)}

Why some groups of redshifts exhibit a phase shift with respect to others is not at present known.

\section{Physical Connections from NGC 3079 to Nearby Quasars}

Fig. 4 is adapted from Pietsch et al. showing PSPC X-ray contours overlayed in white on a PSS 
survey image. Neutral hydrogen, HI, isophotes at the redshift of NGC 3079 are shown in black. 
Redshifts of 5 X-ray quasars are labeled and one BSO candidate are labeled. 

There are 5 and probably more quasars in this small area  and three of them show evidence for 
having interacted with ejected material along the minor axis of NGC 3079 :  
                     
1) The BSO candidate in the center of Fig. 4 is about 37" North of the listed position of the X-ray 
source and has
magnitudes (r = 19.5, b = 19.6). Its offset from the X-ray position is slightly larger than that
which is usually accepted for optical identifications. But the X-ray source is weak and extends to the 
North where the candidate lies, so it seems a likely candidate to be a quasar. The important aspect 
of this possible identification, however, is that the X-ray filament, in which this a thickening, leads 
directly back W , along the path of the previously discusssed z = 1.038 quasar, to the active inner 
regions of the Seyfert galaxy.

2) The HI isophotes around NGC 3073 (Mrk 231) are swept back to SW almost exactly on a line
back to the nucleus of NGC 3079.  It is dificult to avoid the conclusion that a wind or ejection from the
nucleus has ablated the hydrogen gas from this compact Markarian galaxy as concluded by Irwin et 
al. (1987). Furthermore these hydrogen isophotes end directly on the quasar with z = .215. It is of 
course precedented for the Markarian galaxy to have ejected the quasar. But in this particulare case 
it is perhaps more likely that the quasar was ejected fron NGC3079 and that it was carried out along 
with the same wind that is stripping the hydrogen from the Markarian galaxy. Such a mechanism is 
suggested by the third quasar of z = 1.154 just to the NNE of the NGC 3079 disk which we now 
discuss.

3) We quote from the 2 NED references to the quasar of z = 1.154: " The QSO UB4 0957+558 is one of 
three UV-excess QSOs discovered near the galaxies NGC 3079 and Mrk 131 (NGC 3073) by 
Arp (1981). They were identified, as part of a larger sample of QSOs near galaxies, to study the 
statistical association between low-redshift galaxies and quasars." "A number of investigations of 
NGC 3079 several years after Arp's work have shown that it is an exceedingly active galaxy with
extensive evidence of radio emission and a jet-like out-flow from its nucleus . . ." "Of these three 
QSOs, UB4 is the one nearest NGC 3079; . . . we will refer to this object by its accurate . . . J2000 
coordinate name 1002+5542."

The newest evidence is from the ROSAT PSPC maps of NGC 3079 publlshed by Pietsch et al. (1998).
As Fig. 4 here shows there is X-ray material extending from NGC 3079 out to, and enveloping, the
quasar at z = 1.154. Fig. 5, also adapted from Pietsch et al. (1998), shows a full wavelength band,
PSPC  X-ray map of NGC 3079. This is the most convincing data, where one can see a number 
of X-ray contours reaching from the nucleus out directly to the NE square within which the X-ray 
source (Pietsch et al. no. P 21) and the quasar UB4 (z = 1.154) reside. 

There arises the question of whether the optically bright quasar (r = 19.0, b = 19.6) is the correct
identification. Its optical position is 11" N of the X-ray position. Normally accurate enough for an
identification. This is a complicated region, however and higher resolution X-ray maps would be
important to obtain. 

With an X-ray source this strong and this close to a galaxy it would fulfill the definition of a ULX 
(as some of the RIXOS sources also do in this field depending on the admission criterion of projected 
distance).  If we count just U4 as a ULX, however, along with the new ULX which is 8" distant
from the nucleus of NGC 7319 (Galianni 2004, Galianni et al. 2005), then along with the 24 ULX's found 
spectroscopically to be quasars by Arp et al. 2004) the score would now stand: 

\vskip0.5cm

\centerline {Black Hole Binaries =  0}

\centerline{~~~     Quasars~~~~~~~~~~~            = 26}

The importance of the ULX's is, of course, that they have been accepted as physical members of the
their associated galaxies.

If we accept this evidence the question arises as to whether  we can reconcile the 
observations with acceptable physical models?

\section{Relation of the Double Quasar, z = 1.414, to NGC 3079}

The optical images of NGC 3079 show that the NNW end of the galaxy ends with an apparently
straight line of emission knots  extending out to the NNW. The neutral hydrogen map in Fig. 4, 
however, reaches limits which are about twice as large as the optical image. The HI contours at the 
NNW  end become narrow and point closely to p.a. = 340 deg.  As Fig 4 shows, this is closely in 
the direction of the double quasar with z = 1.414. 

There are many radio continuum investigations around this close pair of X-ray bright quasars. The
one covering 40" in declination at 20 cm (Harvanek et al 1997) shows diffuse material and extended 
patches along the p.a. = 340deg. direction which reach back toward NGC 3079. It would possibly be of 
critical importance to measure deeper neutral hydrogen contours at the redshift of NGC 3079 for about
10' between it and the double quasar. The present contours suggest a shallow gradient of the HI
extension from the NW end of NGC 3079 that may reach further with fainter contours.
It would seem to be necessary to get deeper observations in order to disprove the implication that an 
HI bridge leads from NGC 3079 toward the double quasar. 

\section{Ablation from Ejected Bodies?}

The western member of the pair discussed earlier, z = 1.038, is shown in Fig. 6. It has a strong nvss 
radio source just 22" East of the quasar (+ sign). This is directly on a line back to the center of 
NGC 3079. It strongly suggests that the quasar travelling along this path through the ambient 
medium has either encountered a cloud in the halo of NGC 3079 or had a burst of radio emission at this 
point. Since the radio plasma is low density it is then stripped away and we are seeing here the 
beginning of that process. In order make sure of this critical geometry we have measured the 
displacement of the optical position of the quasar with respect to the high resolution (FIRST) radio 
image. The displacement of the quasar to the W is confirmed at 20.4".

In fact we can argue from Fig. 6 that there is both elongation of the core of the radio source and a tail
in the same direction in the two weakest isophotes. That this distortion of the radio source points so
directly to the nearby active nucleus is evidence that the material originated from that nucleus.
The compact quasar appears then to be moving on ahead of this radio source in the direcion of ejection.

This is not the first example of radio plasma being stripped from ejected quasars. Similar
examples of displaced radio sources have been shown near the ejecting Mrk 231, 3C212 and Arp 220 
(see Arp 2001, Figs.9, 10 and 14). Moreover this mechanism can be used to explain the puzzling fact 
that there are relatively few radio sources over the sky which are optically identified whereas almost 
all physically denser X-ray sources are optically identified, usually as quasars. To our knowledge this 
obvious problem has never been seriously addressed.

The other member of the pair at z = .673 has some weak, diffuse nvss material in the NNE direction,
ahead of its presumed flight path. As if the quasar was running into a low density cloud but the data 
seems insufficient to make any conclusions. In general the quasars are overly abundant around 
NGC 3079 in many directions and it would be necessary to conclude that they have come out in a 
number of different directions and that consequently there is a particularly dense extragalactic medium 
in the vicinity of NGC 3079.

If it is clear from the accumulated evidence that NGC 3079 is ejecting optical, X-ray and radio
material then apparently quasars also are being ejected. This is not surprising since quasars radiate
strongly in all three of these wavelength regions - they are only more compact than ejected gas. One 
has to stop and question how is it possible for the radio and and diffuse optical and X-ray plasma to
penetrate the relatively dense medium in the central regions of the ejecting galaxies as well as the 
residue of previous ejecta in the immediate surroundings of the galaxy. One obvious answer is that 
the initial ejecta are relatively compact objects which can burrow through the ambient medium. They 
can then entrain or hollow a channel through the medium through which galaxy gas flows as in a 
"wind" (or ablate their less compact outer regions into their trails).

Going back to the closest quasar to NGC 3079, the z = 1.154 object, known originally as U4, we see 
In Figs 5 and 6 that the ejection of material NE from the NGC 3079 nucleus might contain bright
and faint quasars, X-ray concentrations and diffuse gas. If the quasar is ablating it gives us
for the first time the opportunity to learn something about its physical nature. 
The low particle mass plasma which is suggested to account for the intrinsic redshift of the
quasar (Narlikar and Arp 1993) may be dense enough and small enough to make a path through the 
medium but fragile enough to break up or ablate. Deeper X-ray observations and HST
image analysis would be quite important in this region. Pietsch et al. have XMM observations in 
preparation which may give a much clearer picture of the events in the region of this z=1.154 quasar.

\section{Summary}

The nine X-ray brightest quasars which are conspicuously grouped around the active Seyfert
NGC 3079 have been tested for statistical probability of association. The resulting probabilities 
range from $10^{-3} to 10^{-4}$ of being accidental. Computation from expected variations of 
background density gives about a 10 sigma significance. Allowing for the large overestimate
of current X-ray/AGN background densities would make all these associations even more significant.

Accepting the statistical evidence for association, the individual quasars can then be investigated for
interaction with the material in the vicinity of NGC 3079. A picture emerges where relatively
compact quasars are ejected along with X-ray and radio material. Low density radio and possibly
some X-ray material can be subsequently ablated by the medium through which they pass. The
process of radio plasma stripping is suggested as an explanation of the generally ignored  problem 
that many radio sources in the sky are optically unidentified whereas most X-ray sources are 
measured to be quasar/AGN's.

As for the much studied double quasar only 14' from NGC 3079, two pertinent data are reported:

1) The X-ray bright quasars, including the double quasar, within this area around NGC 3079 are overly 
dense by about 10 sigma from a conservative estimate of the all sky density of such objects.

2) HI extensions from NGC 3079 point rather closely in the direction of the double quasar and there
is some alignment of continuum sources around the double quasar in this same direction.

The suggestion would be that the double quasar, like the rest of the X-ray bright quasars around NGC
3079 and similarly active galaxies, has a non-velocity, intrinsic redhift.

\clearpage

\section{REFERENCES}

Arp, H. 1981, ApJ 250, 31

Arp, H. 1997, A\&A, 319, 33

Arp, H. 1998, Seeing Red, Apeiron, Montreal

Arp, H. 1999 A\&A, 341 L5 

Arp, H. 2001 ApJ, 549, 780.

Arp, H. 2003, Catalogue of Discordant Redshift Associations, Apeiron, Montreal

Arp, H., Bi, H., Chu, Y., Zhu, X. 1990 A\&A 239, 33

Arp, H., L\'opez-Corredoira, M. \& Guti\'errez, C. M. 2004, A\&A, 418, 877

Burbidge, G., Burbidge, E. M. \& Arp, H. 2003, A\&A, 400, L17

Burbidge, E. M., Guti\'errez, C. \& Arp, H. 2004  A\&A submitted

Burbidge, E. M.; Burbidge, G.; Arp, H.; Zibetti, S. 2004, ApJS 153,159

Chu, Y., Wei, J. Hu, J., Zhu, X, Arp, H. 1998 ApJ, 500, 596

Colbert, E. \& Ptak A. 2002, ApJS, 143, 25

Galianni, P. 2004, Coelum Feb 2004, 54

Galianni, P.. Burbidge, E. M., Arp, H. Junkkarinen V., Zibetti, S.  2005, ApJ, Feb. in press

Harvanek, M., Stocke, J., Morse, J., Rhee, G. 1997, AJ....114, 2240

Hasinger, G.; Burg, R.; Giacconi, R.,Hartner, G., Schmidt, M., Trumper, J. Zamorani, G.1993, 
A\&A 275, 1

Hasinger, G.; Burg, R.; Giacconi, R.; Schmidt, M.; Trumper, J.; Zamorani, G.,1998, A\&A, 329, 482

Irwin, Judith A.; Seaquist, E. R. Taylor, A. R.; Duric, N. 1987, ApJ 13L,91

Page, M. J.; Carrera, F. J.; Hasinger, G.; Mason, K. O.; McMahon, R. G.; Mittaz, J. P. D.; Barcons, X.; 
Carballo, R.; Gonzalez-Serrano, I.; Perez-Fournon, I. 1996MNRAS.281..579P

Pietsch, W., Trinchierii, G., Vogler, A. 1998, A\&A 340, 351

Pietsch, W., Arp, H. 2001, A\&A, 376, 393

Radecke, H.-D. 1996, Max-Planck-Institut f\"ur Astrophysik, preprint MPA 965 June 1996

Radecke, H.-D. 1997, A\&A, 319, 18

Sandage, A., Tamman, G. 1981, A Revised Shapley-Ames Catalog, Carnegie Institution of Washington,
1981

Stepanian, J., Lipovetsky, V., Erastova, L. 1990, Astrofizika 23, 441

\clearpage

Fig. 1  ---  The brightest X-ray sources within 50' radius of the disturbed , ejecting Seyfert galaxy 
NGC 3079. Redshifts written next to the AGN's.

Fig. 2  ---  The brightest X-ray sources within 15' radius of NGC 3079, Image of galaxy inserted to 
scale. 

Fig. 3  -- The high resolution FIRST radio map showing radio ejection coming out in opposite directions
from the nucleus.

Fig. 4  --  A red PSS survey photograph overlayed with white contours for X-rays and black countours
for neutral hyrogen. Adapted from Pietsch et al 1998 which was adapted from Irwin et al 1987.
Redshifts of X-ray sources are labeled.

Fig. 5  -- ROSAT broad band PSPC X-ray contours for NGC 3079. Square marks region within which
the X-ray point source P 21 is located (identified here as quasar z = 1.154). Adapted from 
Pietsch et al. 1998. 

Fig. 6  -- The nvss radio map of the z = 1.038 quasar W of NGC 3079. Notice the $\sim$ 21" offset 
of the quasar (+ sign) to the W of the radio source. Also the elongation of the innermost and outermost
radio contours.

\clearpage

\begin{figure}[h]
\includegraphics[width=16.0cm]{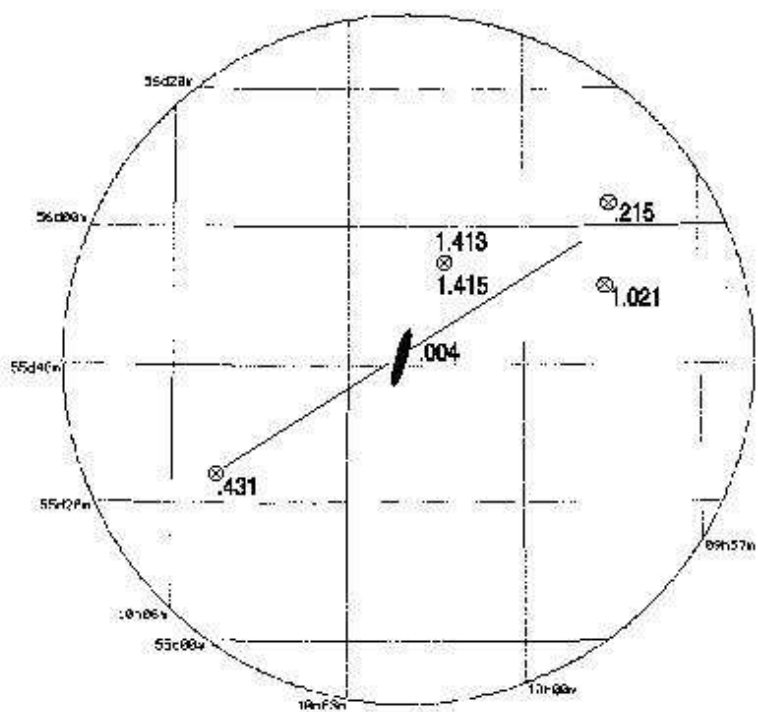}
\caption{The brightest X-ray sources within 50' radius of the disturbed, ejecting Seyfert galaxy 
NGC 3079. Redshifts written next to the AGN's.
\label{fig1}}
\end{figure}

\clearpage

\begin{figure}
\includegraphics[width=16.0cm]{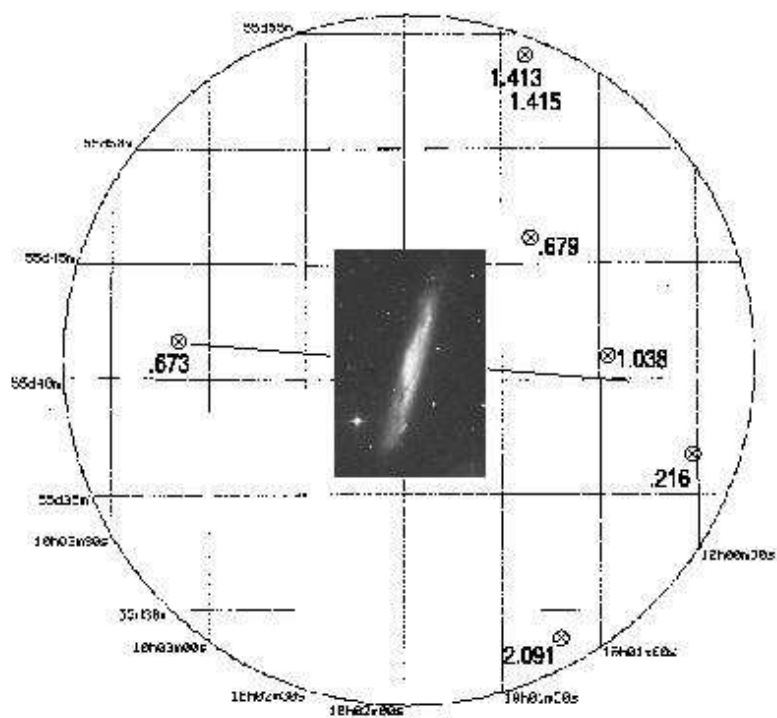}
\caption{The brightest X-ray sources within 15' radius of NGC 3079, Image of galaxy inserted to 
scale. 
\label{fig2}}
\end{figure}

\clearpage

\begin{figure}
\includegraphics[width=12cm]{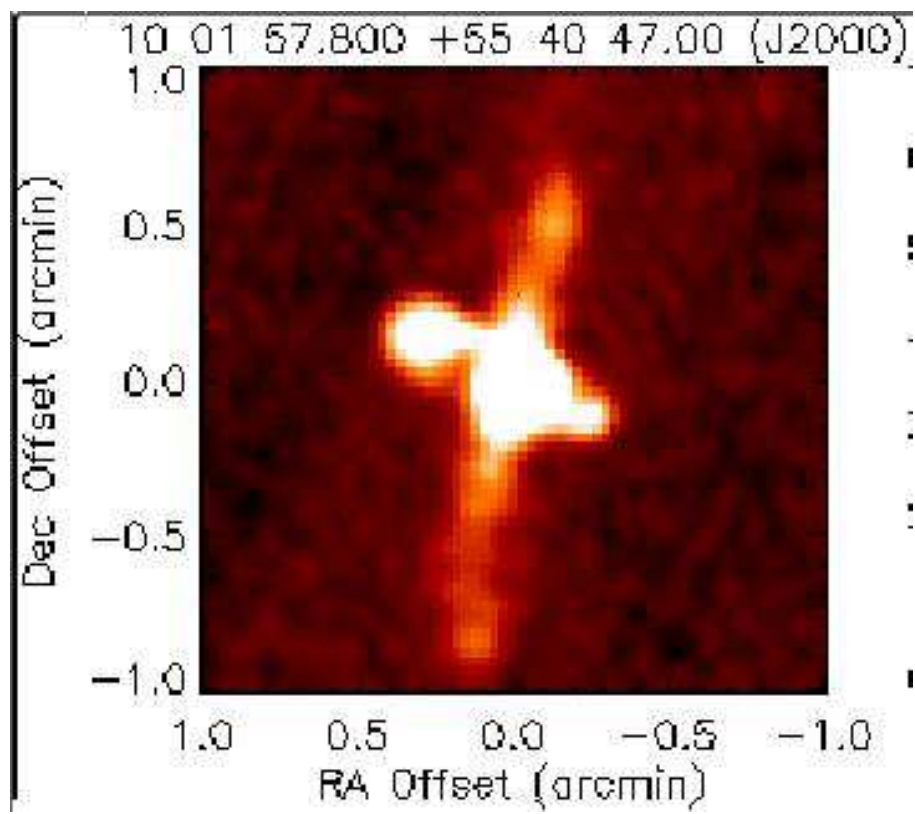}
\caption{The high resolution FIRST radio map showing radio ejection coming out in opposite directions
from the nucleus. \label{fig3}}
\end{figure}

\begin{figure}
\includegraphics[width=16cm]{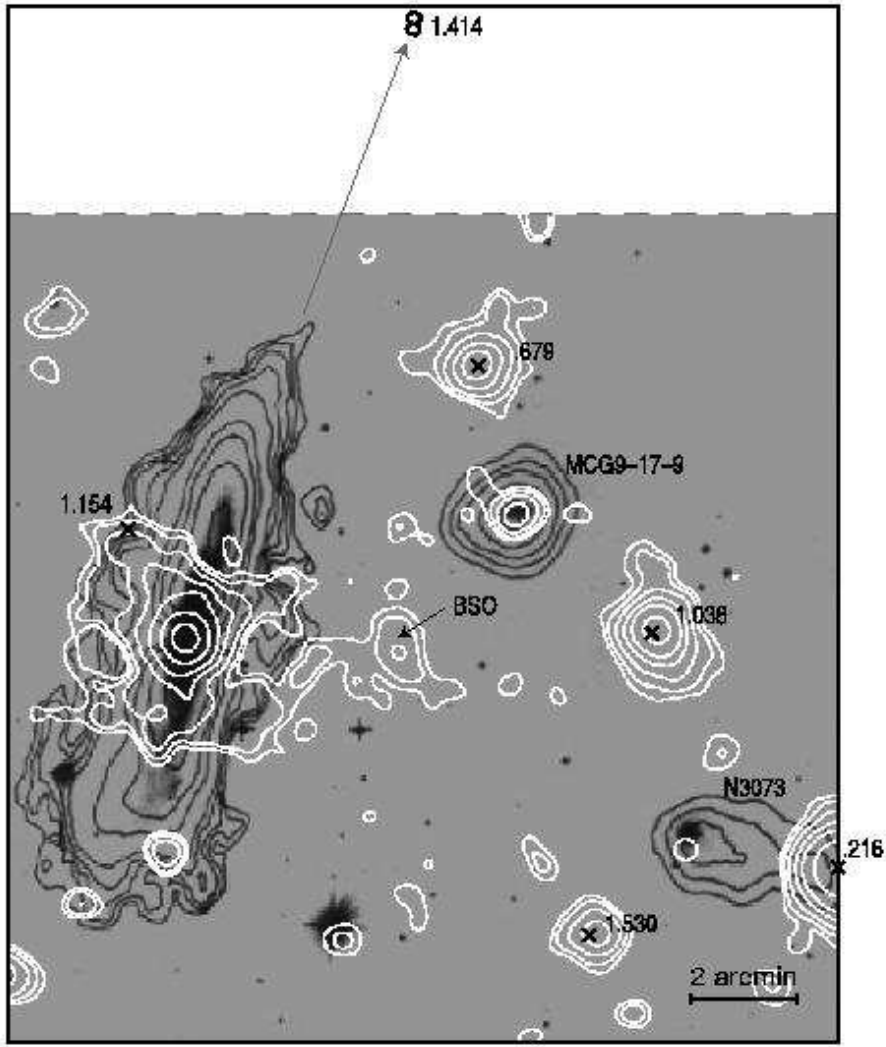}
\caption{A red PSS survey photograph overlayed with white contours for X-rays and black countours
for neutral hyrogen. Adapted from Pietsch et al 1998 which was adapted from Irwin et al 1987.
Redshifts of X-ray sources are labeled. \label{fig4}}
\end{figure}

\clearpage

\begin{figure}
\includegraphics[width=14cm]{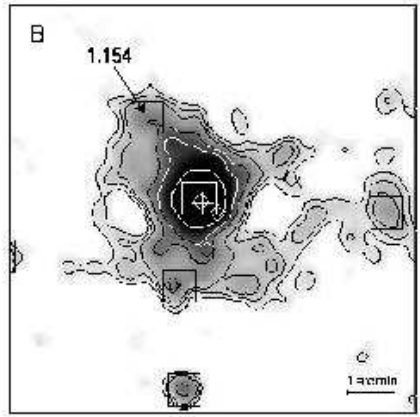}
\caption{ROSAT broad band PSPC X-ray contours for NGC 3079. Square marks region within which
the X-ray point source P 21 is located (identified here as quasar z = 1.154). Adapted from 
Pietsch et al. 1998. \label{fig5}}
\end{figure}

\clearpage

\begin{figure}
\includegraphics[width=18cm]{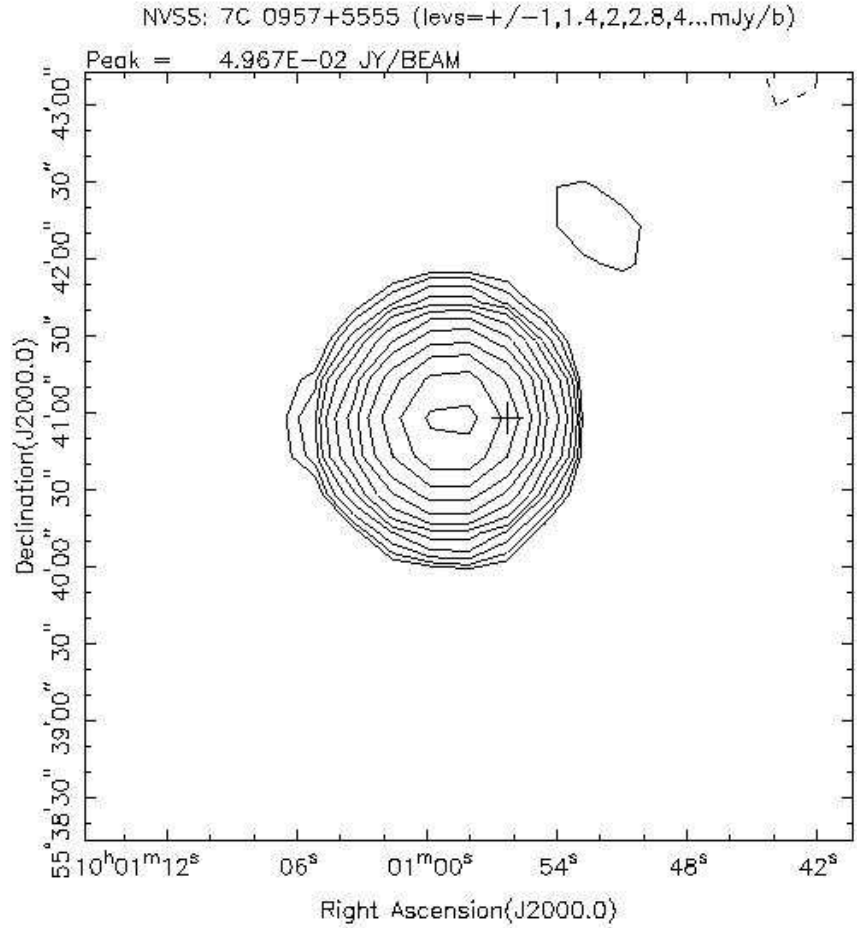}
\caption{The nvss radio map of the z = 1.038 quasar W of NGC 3079. Notice the $\sim$ 21" offset 
of the quasar (+ sign) to the W of the radio source. Also the elongation of the innermost and 
outermost radio contours. 
\label{fig6}}
\end{figure}

\end{document}